\documentclass{moriond}
\usepackage[english]{babel}
\usepackage{graphicx}
\usepackage{graphics}
\usepackage{braket}
\usepackage{bbold}
\usepackage{amssymb}
\usepackage{amsmath}
\usepackage{nicefrac}
\usepackage{dcolumn}% Align table columns on decimal point
\usepackage{bm}% bold math
\usepackage{slashed}
\usepackage{datetime}
\usepackage{mciteplus}
\usepackage{multirow}
\usepackage{bbold}
\usepackage{siunitx}
\usepackage{booktabs}
\usepackage{color, soul}
\usepackage[usenames,dvipsnames]{xcolor}
\usepackage{float}
\usepackage[utf8]{inputenc}
\usepackage[normalem]{ulem}
\usepackage{mathtools}

\newcommand{\Jps}{J/\psi}
\newcommand{\Yps}{\Upsilon}

\newcommand{\ebs}{\eta_b}

\newcommand{{\HFNRevo}}{\tt HF-NRevo}

\def\be{\begin{equation}}
\def\ee{\end{equation}}
\def\bea{\begin{eqnarray}}
\def\eea{\end{eqnarray}}

\newcommand{\Photo}{\includegraphics[height=35mm]{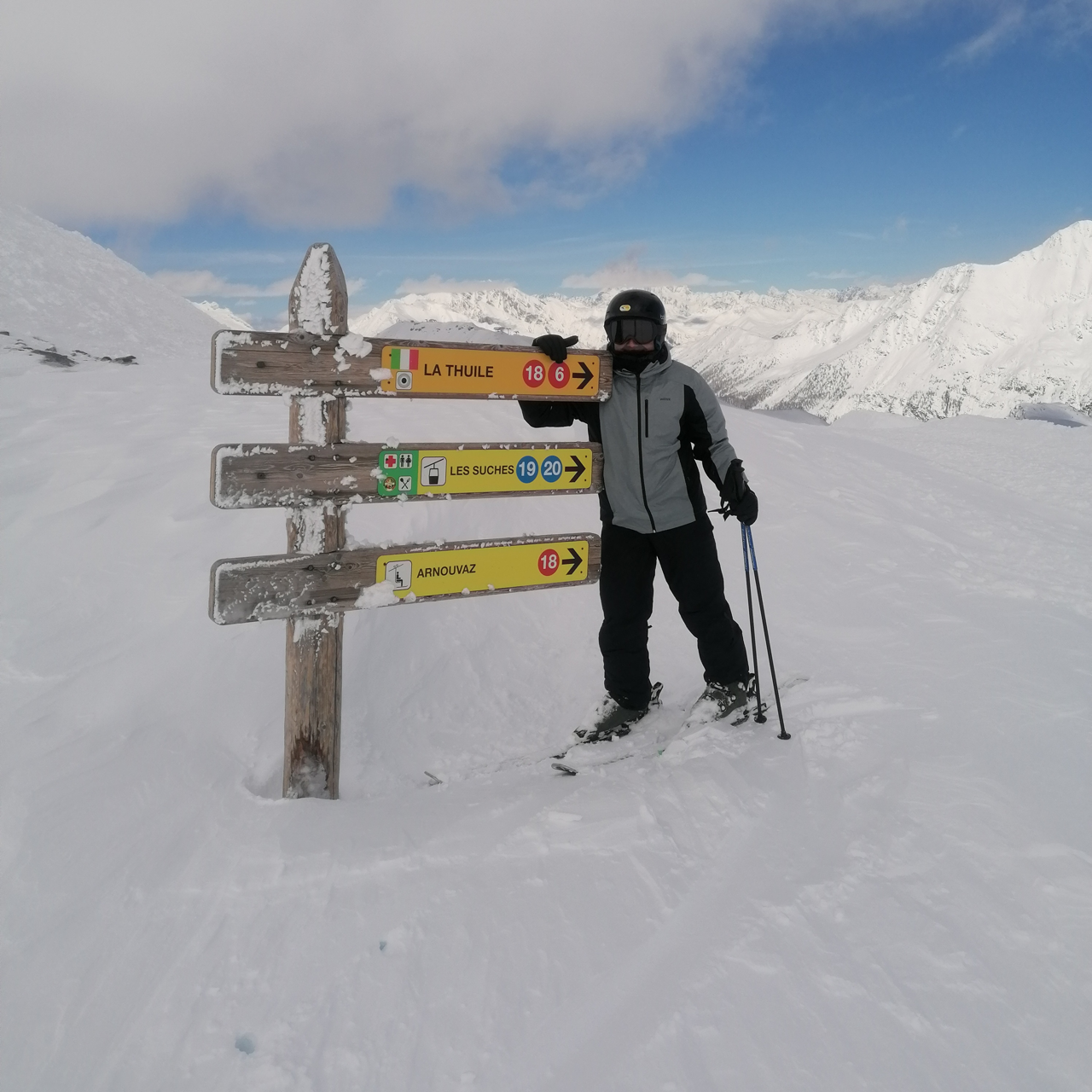}}

\begin{document}
\vspace*{4cm}
\title{Towards Quarkonium Fragmentation from NRQCD \\ in a Variable-Flavor Number Scheme}

\author{Francesco Giovanni Celiberto$^{\;1\;}$\footnote{%Corresponding Author. 
Electronic address: \href{mailto:francesco.celiberto@uah.es}{francesco.celiberto@uah.es}}}

\address{
${}^1$Universidad de Alcal\'a (UAH), E-28805 Alcal\'a de Henares, Madrid, Spain
}

\maketitle\abstracts{
We address  quarkonium formation at moderate to large transverse mo\-men\-ta, where the single-parton collinear fragmentation prevails over the short-distance emission, directly from the hard sub-scattering, of the constituent heavy-quark pair.
We rely on Non-Relativistic-QCD (NRQCD) Next-to-Leading Order (NLO) calculations for all parton fragmentation channels to quarkonia, taken as proxies for initial-scale inputs. 
Preliminary sets of Variable-Flavor Number-Scheme (VFNS) fragmentation functions (FFs) are built via a DGLAP scheme that properly accounts for evolution thresholds.
Statistical errors are assessed via a Monte Carlo (MC), replica-like approach aimed at catching Missing Higher-Order Uncertainties (MHOUs).
}

\section{Hors d'{\oe}uvre}
\label{sec:introduction}

%If you more commonly use the method of square brackets in the line of text
%for citation than the superscript method,
%please note that you need  to adjust the punctuation
%so that the citation command appears after the punctuation mark.

Within the realm of fundamental interactions, hadrons with open as well as hidden heavy flavor(s) hold a special place.
Heavy quarks are sentinels in the quest for signals of New Physics, given their expected couplings with BSM objects. 
Being their masses above the perturbative QCD threshold, they also allow for stringent tests of the strong force.
Among them, quarkonia stand as building blocks for several research directions.
They can be imagined as ``hydrogen atoms'' of QCD.~\cite{Pineda:2011dg}
Quarkonium physics resides at the intersection of two complementary aspects of frontier research: \emph{precision} studies employing perturbative techniques on one hand, and \emph{explorations} of the hadronic structure on the other.
Hadronic inclusive decays of $S$-wave bottomonia enable precision determinations of the QCD running coupling.~\cite{Brambilla:2007cz,Proceedings:2019pra} 
Forward inclusive quarkonium detections can test the positivity of gluon parton densities (PDFs) at low-$x$ and low-$Q^2$.~\cite{Altarelli:1998gn}$^-$\cite{Candido:2023ujx}
Quarkonia are excellent tools for a 3D tomographic imaging of the proton at low~\cite{Bautista:2016xnp}$^-$\cite{Kang:2023doo} and moderate $x$.~\cite{Boer:2015pni}$^-$\cite{Celiberto:2021zww} 
Notably, the unresolved photoproduction of a $\Jps$ plus a tagged charm-jet at the EIC~\cite{Flore:2020jau} will allow us to assess the weight of the intrinsic-charm~\cite{Brodsky:1980pb}$^-$\cite{Guzzi:2022rca} valence PDF in the proton.~\cite{NNPDF:2023tyk} 
%%%%%%
Despite being experimentally accessible, the theoretical depiction of quarkonium hadron\-ization remains challenging. 
Various models have been proposed, yet none comprehensively capturing all experimental signatures.~\cite{Grinstein:1998xb}$^-$\cite{Lansberg:2019adr} 
To unveil the quarkonium production puzzle, an effective theory, known as NRQCD, was formulated.~\cite{Caswell:1985ui}$^-$\cite{Bodwin:1994jh}  
NRQCD assumes that all possible Fock states contribute to the physical quarkonium.
They are systematically organized into a double expansion in powers of the strong coupling, $\alpha_s$, and the relative velocity of the constituent heavy quarks, $v$.
Quarkonium production rates are expressed as a sum of partonic Short-Distance Coefficients (SDCs), each multiplied by a Long-Distance Matrix Element (LDME) describing the non-perturbative hadronization.
NRQCD permits rigorous tests of quarkonium formation mechanisms, particularly the \emph{short-distance} emission of a $(Q \bar Q)$ pair directly produced in hard scattering, dominant at low transverse momentum, $|\vec p_T|$. 
As $|\vec p_T|$ increases, however, another mechanism gains significance: \emph{fragmentation}, wherein a single parton undergoes inclusive decay into the observed quarkonium.~\cite{Cacciari:1994dr}
%%%%%%
We present preliminary results on new determinations of collinear FFs for color-singlet (CS) vector and pseudoscalar quarkonia, which we name {\tt NRFF1.0} functions.
The rely upon a novel method, known as Heavy-flavor Non-Relativistic Evolution ({\HFNRevo}), which builds on NRQCD initial-scale inputs at NLO and accounts for a proper DGLAP evolution, as well as MHOU-based uncertainties from a MC, replica-like treatment.~\cite{Forte:2002fg} 

\section{Heavy-flavor Non-Relativistic Evolution}
\label{sec:HFNrevo}

\begin{figure*}[!t]
\centering

   \includegraphics[scale=0.38,clip]{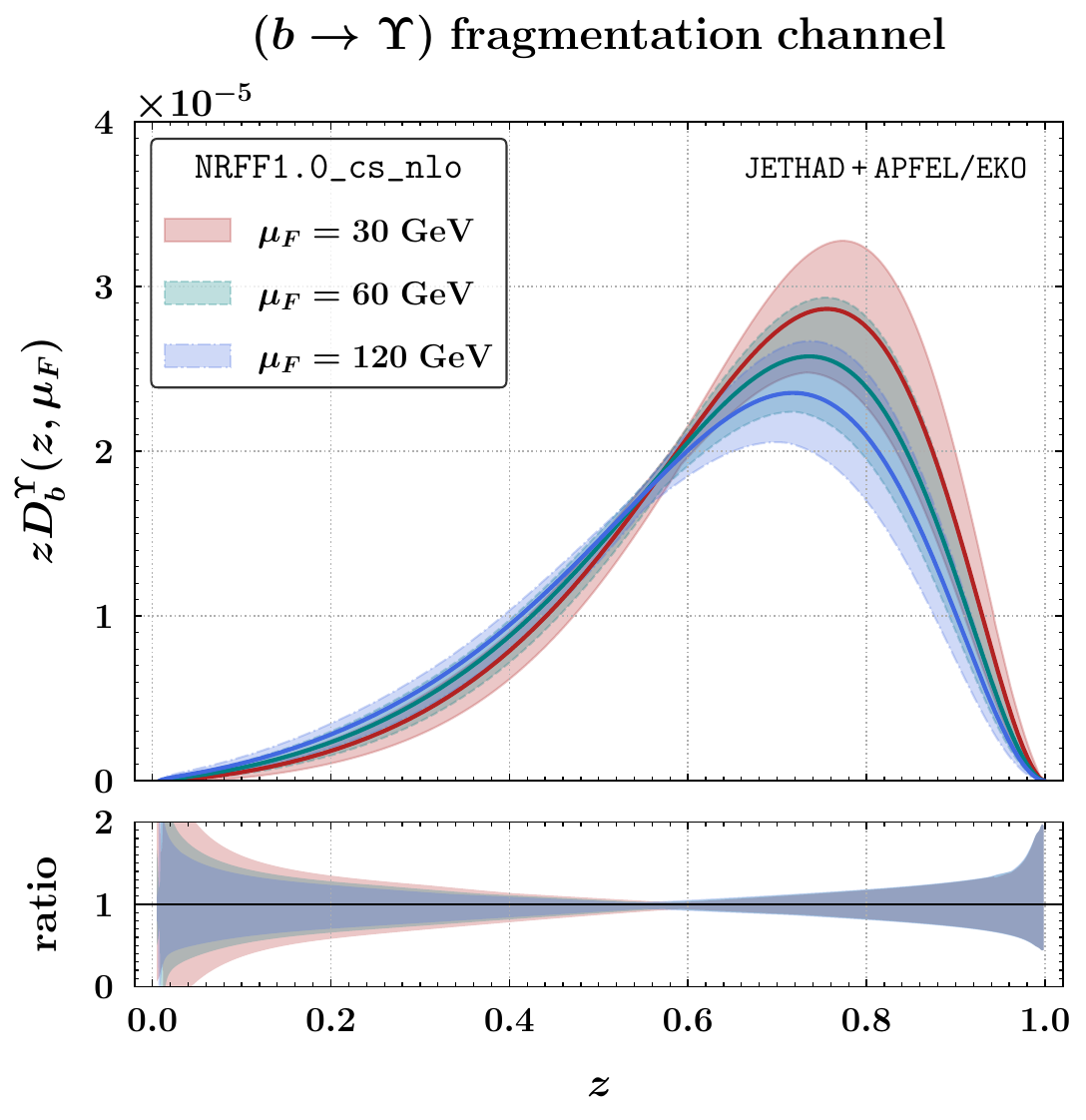}
   \hspace{0.50cm}
   \includegraphics[scale=0.38,clip]{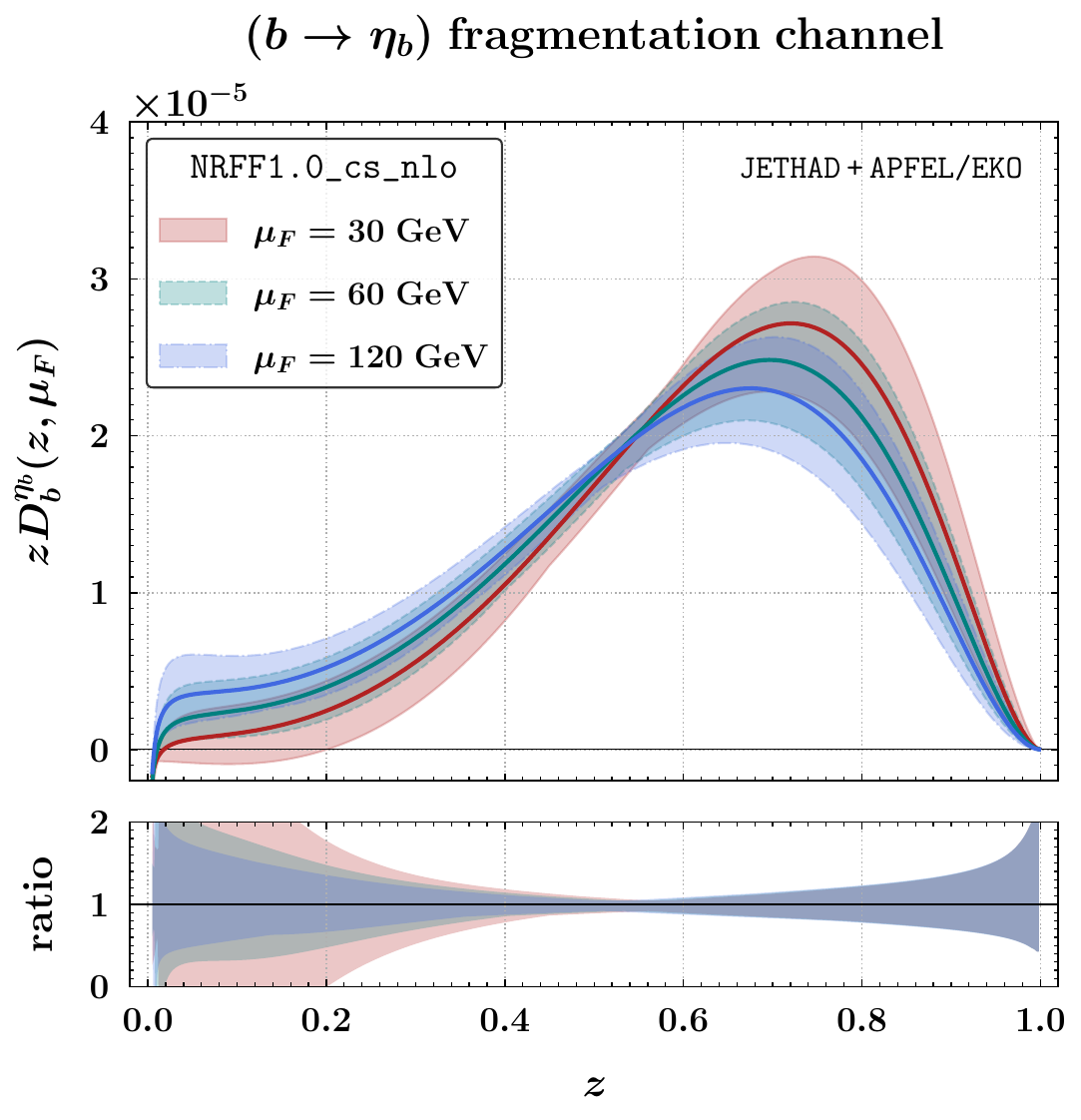}

\caption{Bottom-quark  NLO FF to CS $\Yps$ and $\ebs$ mesons. Preliminary results for {\tt NRFF1.0}.
}

\label{fig:FFs}
\end{figure*}

Since masses of heavy quarks in the lowest Fock state $|Q \bar{Q}\rangle$ of quarkonia are above the per\-turb\-ative-QCD threshold, initial-scale inputs of corresponding FFs are expected to incorporate some perturbative elements.
While NRQCD factorization offers a natural and elegant solution, given by the convolution of SDCs and LDMEs (see~\cite{Mele:1990cw,Cacciari:1993mq} for an analogous treatment of hadrons with open heavy flavor), a proper connection with the collinear factorization is needed.
To this end, we have developed a novel methodology, named {\HFNRevo}.
It builds upon three founding pillars: interpretation, evolution, and uncertainties.
The interpretation allows us decipher the short-distance formation, dominant at low $|\vec p_T|$, as a Fixed-Flavor Number-Scheme (FFNS) two-parton fragmentation.
This sets the stage for a subsequent matching between FFNS and VFNS calculations.~\cite{Kang:2014tta} 
This statement is corroborated by the observation that, when transverse-momentum dependence is accounted for, the matching tails of low-$|\vec p_T|$ shape functions~\cite{Echevarria:2019ynx} and moderate-$|\vec p_T|$ FFs~\cite{Echevarria:2020qjk,Copeland:2023wbu} exhibit different kinds of divergences.~\cite{Boer:2023zit} 
Following {\HFNRevo}, the DGLAP evolution of our quarkonium FFs can be split in two steps.
First, an \emph{expanded} and \emph{decoupled} evolution ({\tt EDevo}, performed with {\tt JETHAD}~\cite{Celiberto:2020wpk}$^-$\cite{Bolognino:2021mrc} symbolically), properly accounts for evolution thresholds of the various parton species.
Then, the standard \emph{all-order} evolution ({\tt AOevo}, performed with~{\tt APFEL++}~\cite{Bertone:2013vaa}  numerically; linking our technology to {\tt EKO}~\cite{Candido:2022tld} is underway) is switched on.
Finally, we assess the weight of MHOUs coming from scale variations connected to DGLAP-evolution thresholds.
This strategy is close in spirit with studies on PDFs, adopting the theory-covariance-matrix approach~\cite{Harland-Lang:2018bxd}$^-$\cite{NNPDF:2024dpb} or the {\tt MCscales} method.~\cite{Kassabov:2022orn} 
%%%%%%
For brevity, we present just two fragmentation channels, \emph{i.e.} constituent bottom to bottomonium FFs.~\cite{Braaten:1993mp}$^-$\cite{Zheng:2021ylc} 
Left and right plots of Fig.~\ref{fig:FFs} show the $z$-behavior of $(b \to \Yps)$ and $(b \to \ebs)$ {\tt NRFF1.0} FFs, for three values of $\mu_F$.

\section{Towards {\tt NRFF1.0} and beyond}
\label{sec:conclusions}

Using the new {\HFNRevo} methodology, we derived a preliminary version of the {\tt NRFF1.0} quarkonium collinear FFs, which incorporate CS initial-scale inputs from all parton channels calculated within NLO NRQCD.
They build upon a consistent DGLAP scheme that properly accounts for evolution thresholds, as well as MHOU-based uncertainties from a MC, replica-like treatment.
The {\tt NRFF1.0} FFs will supersede the {\tt ZCW19}$^+$ and {\tt ZCFW22} sets currently employed in phenomenological studies of vector quarkonia~\cite{Celiberto:2022dyf,Celiberto:2023fzz} and $B_c$ mesons.~\cite{Celiberto:2022keu,Celiberto:2024omj} 
They will serve as a guidance for explorations of quarkonium physics at the HL-LHC,~\cite{Chapon:2020heu,Amoroso:2022eow} the EIC,~\cite{AbdulKhalek:2021gbh}$^-$\cite{Abir:2023fpo} and lepton colliders,~\cite{ILCInternationalDevelopmentTeam:2022izu} as well as a reference for prospective AI-based extractions.~\cite{Allaire:2023fgp} 
Future extensions will include: color-octet studies,~\cite{Cho:1995vh}$^-$\cite{Boussarie:2017oae} a GM-VFNS matching,~\cite{Cacciari:1998it}$^-$\cite{Buza:1996wv} and applications to the exotic matter.~\cite{Celiberto:2023rzw,Celiberto:2024mrq} 

\section*{Acknowledgments}
 
This work is supported by Atracci\'on de Talento Grant n. 2022-T1/TIC-24176, Madrid, Spain.

%\section*{Appendix}

\end{document}